\documentclass[conference]{IEEEtran}
\IEEEoverridecommandlockouts

\usepackage{cite}
\usepackage[square,numbers]{natbib}
\usepackage[mode=buildnew]{standalone}

\usepackage{amsmath,amssymb,amsfonts}
\usepackage{algorithmic}
\usepackage{graphicx}
\usepackage{textcomp}
\usepackage{xcolor}

\usepackage{float}
\usepackage{listings}             
\newfloat{lstfloat}{htbp}{lop}
\floatname{lstfloat}{Listing}

\lstdefinelanguage{predicate}{
    showspaces=false,
    captionpos=b,
    basicstyle=\ttfamily\small,
    upquote=true,
    morestring=[b]",
    stringstyle=\color{purple},
    morekeywords={AND, OR, WHERE}
}

\usepackage{caption}
\usepackage{subcaption}
\usepackage{pgfplots}

\usepackage{booktabs} 
\usepackage{csvsimple}%

\usepackage{tabularx}  
\usepackage{array,multirow}
\usepackage{diagbox}

\usepackage{tikz}

\usetikzlibrary{calc,positioning,shapes,intersections,backgrounds,scopes,fit,arrows.meta}
\usetikzlibrary{shapes.geometric, tikzmark}

\tikzset{
  invisible/.style={opacity=0},
  visible on/.style={alt=#1{}{invisible}},
  alt/.code args={<#1>#2#3}{%
    \alt<#1>{\pgfkeysalso{#2}}{\pgfkeysalso{#3}} 
  },
}

\tikzstyle{pipelem} =[fill=gray!60, draw=none]

\definecolor{hwback}{RGB}{224,243,219}
\definecolor{hwcontrast}{RGB}{168,221,181}
\definecolor{hwfront}{RGB}{67,162,202}
\definecolor{hwborder}{RGB}{8,104,172}

\definecolor{fpga}{RGB}{110, 212, 255}
\definecolor{cpu}{RGB}{202, 139, 67}

\definecolor{mygreen}{RGB}{0,128,128}
\definecolor{myorange}{RGB}{255,128,0}
\definecolor{myred}{RGB}{255,76,76}

\definecolor{mygray}{RGB}{220,220,220}
\definecolor{myyellow}{RGB}{255,255,160}

\definecolor{myyellow}{RGB}{255,255,160}

\definecolor{bordercolor}{RGB}{55,55,60}

\definecolor{cbrew1}{RGB}{141,211,199}
\definecolor{cbrew2}{RGB}{255,255,179}
\definecolor{cbrew3}{RGB}{190,186,218}
\definecolor{cbrew4}{RGB}{251,128,114}
\definecolor{cbrew5}{RGB}{128,177,211}
\definecolor{cbrew6}{RGB}{253,180,98}
\definecolor{cbrew7}{RGB}{179,222,105}
\definecolor{cbrew8}{RGB}{252,205,229}
\definecolor{cbrew9}{RGB}{217,217,217}

\colorlet{cloudcolor}{cbrew5}

\colorlet{acccolor}{hwback}

\colorlet{buscolor}{hwcontrast}

\colorlet{prcolor}{fpga}
\colorlet{staticcolor}{cbrew4}
\colorlet{cpucolor}{cpu}

\colorlet{corecolor}{mygray}
\colorlet{gearcolor}{mygray}

\colorlet{ssdcolor}{myorange}
\colorlet{nvramcolor}{myorange!70!white}
\colorlet{ramcolor}{myorange!50!white}

\definecolor{shade1}{RGB}{110, 212, 255}
\definecolor{shade2}{RGB}{29, 97, 125}
\definecolor{shade3}{RGB}{67, 162, 202}

\definecolor{coldesign}{RGB}{0, 0, 0}

\definecolor{myblue}{RGB}{55,126,184}
\definecolor{mypurple}{RGB}{152,78,163}

\definecolor{bordercolor}{RGB}{0,0,0}

\definecolor{reocolor}{RGB}{141,211,199}
\definecolor{pscolor}{RGB}{255,255,179}
\definecolor{memcolor}{RGB}{251,128,114}
\definecolor{bgcolor}{RGB}{217,217,217}

\definecolor{color1}{RGB}{127,201,127}
\definecolor{color3}{RGB}{190,174,212}
\definecolor{color2}{RGB}{253,192,134}
\definecolor{color4}{RGB}{255,255,153}

\colorlet{datacolor}{ssdcolor!80}
\colorlet{softwarecolor}{corecolor!80}
\colorlet{operatorcolor}{prcolor!80}
\colorlet{inputcolor}{datacolor!30}

\tikzstyle{innerWhite} = [semithick, white,line width=1.4pt, shorten >= 4.5pt]

\usepackage{tikz/circuitikzgit}
\usepackage{pgfmath}

\usepackage[acronym,toc]{glossaries}
\newacronym{b}{B}{block length}
\newacronym{fpr}{FPR}{false-positive rate}
\newacronym{senml}{SenML}{Sensor Measurement Lists}
\newacronym{ps}{PS}{processor system}
\newacronym{pl}{PL}{programmable logic}
\newacronym{rf}{RF}{Raw Filter}
\newacronym{dfa}{DFA}{deterministic finite automata}

\def\BibTeX{{\rm B\kern-.05em{\sc i\kern-.025em b}\kern-.08em
    T\kern-.1667em\lower.7ex\hbox{E}\kern-.125emX}}

\usepackage[hidelinks]{hyperref}

\usepackage{cleveref}
\Crefname{lstlisting}{Listing}{Listings}

\newlength\bibitemsep
\usepackage[
    all=normal,
    floats=tight,
    leading=tight,
    leadingfraction=0.93,
    paragraphs=tight,
    charwidths=tight, 
    charwidthfraction=0.93,
    mathdisplays=normal,
    tracking=normal,
    lists=normal,
    bibnotes=tight,
    wordspacing=normal
]{savetrees}

\begin{document}

\title{Raw Filtering of JSON Data on FPGAs}

\author{\IEEEauthorblockN{Tobias Hahn, Andreas Becher, Stefan Wildermann, J\"urgen Teich}
    \IEEEauthorblockA{\textit{Chair of Hardware/Software Co-Design, Friedrich-Alexander-Universit\"at Erlangen-N\"urnberg (FAU), Germany} \\
        $<$tobias.hahn,andreas.becher,stefan.wildermann,juergen.teich$>$@fau.de}
}

\maketitle

\begin{abstract}

    Many Big Data applications include the processing of data streams on semi-structured data formats such as JSON.
    A disadvantage of such formats is that an application may spend a significant amount of processing time just on unselectively parsing all data.
    To relax this issue, the concept of raw filtering is proposed with the idea to remove data from a stream prior to the costly parsing stage.
    However, as accurate filtering of raw data is often only possible after the data has been parsed, raw filters are designed to be approximate in the sense of
    allowing false-positives in order to be implemented efficiently.
    
    Contrary to previously proposed CPU-based raw filtering techniques that are restricted to string matching, we present
    FPGA-based primitives for filtering strings, numbers and also number ranges. 
    In addition, a primitive respecting the basic structure of JSON data is proposed that can be used to further increase the accuracy of introduced raw filters.
    
    The proposed raw filter primitives are designed to allow for their composition according to a given filter expression of a query. 
    Thus, complex raw filters can be created for FPGAs which enable a drastical decrease in the amount of generated false-positives, particularly for IoT workload. 

    As there exists a trade-off between accuracy and resource consumption, we evaluate primitives as well as composed raw filters using different queries from the RiotBench benchmark.
    Our results show that up to 94.3\% of the raw data can be filtered without producing any observed false-positives using only a few hundred LUTs.
\end{abstract}


\section{Introduction}

Many Big Data applications in domains such as the Internet of Things and Industry 4.0 not only face a high volume of data that is generated with a high frequency, but also have strong latency requirements for analyzing this data.
In this realm, data stream processing is gaining more and more importance where data is continuously processed and analyzed as soon as it is generated or received.
However, data is often streamed in unstructured or semi-structured data formats, such as JSON, XML, SenML. 
While these formats are easily readable for humans, it is in fact quite the opposite for CPUs: 
Parsers have to process ASCII characters byte by byte, not fully utilizing the CPU word width. 
\citet{mison} reports that Big Data applications can spend 80\%-90\% of their execution time on parsing.  
Our own evaluations of the Yahoo streaming benchmark\footnote{https://github.com/yahoo/streaming-benchmarks} have shown that parsing takes 70\% of the processing time.

%
\begin{lstlisting}[
    float=tp, 
    language=predicate, 
    caption={Running Example - JSON record excerpt taken from the RiotBench SmartCity dataset.},
    label=lst:running_exp_rec,
    belowskip=-0.3cm
    ]
{"e":[ ...
    {"v":"35.2","u":"far","n":"temperature"},
    {"v":"12","u":"per","n":"humidity"},
    {"v":"713","u":"per","n":"light"},
    {"v":"305.01","u":"per","n":"dust"},
    {"v":"20","u":"per","n":"airquality_raw"}
],"bt":1422748800000}
\end{lstlisting}
\begin{lstlisting}[
    float=tp, 
    basicstyle=\normalfont\ttfamily,
    language=predicate,
    captionpos=b,
    caption=Running Example - JSONPath Query,
    escapeinside={(*}{*)},        label=lst:running_exp_q,
    belowskip=-0.3cm
    ]
(*$Q_0$*) := $.e[?(@.n=="temperature" 
            & @.v (*$\geq$*) 0.7 & @.v (*$\leq$*) 35.1)]
\end{lstlisting}
Modern parser implementations, like Mison \cite{mison} and simdjson \cite{simd-json} for JSON, utilize SIMD instructions to mitigate this problem and can scan documents with up to 2 GB/s with one core.
However, sequentially scanning a raw byte stream with a single core would still be $10\times$ faster than retrieving the (JSON) structure by parsing it \cite{sparser}.
Therefore, \citet{sparser} suggested raw filtering of the byte stream \emph{before} the parser is applied with the idea of exploiting the selectivity of the given query to further accelerate JSON parsing. 
Filter predicates are translated into so-called \glspl{rf}, which are inspecting the raw byte stream in a structure-agnostic fashion. 
\Cref{lst:running_exp_q} presents a query on a stream of IoT records as illustrated in \Cref{lst:running_exp_rec} as a motivational example. 
Here, all records containing a "temperature" sensor measure which lies in the range of $[0.7, 35.1]$ are queried. 
\glspl{rf} proposed by \cite{sparser} are restricted to string comparisons. 
The "temperature" string would thus be a possible \gls{rf} to be searched in the byte stream regardless of any structure. 
If the string is found at least once in a record, the record is accepted. 
If no occurrences of the string can be identified, the record is dropped and the parser doesn't have to deal with it. 
While this approach may produce some \textit{false-positives} (e.g., records where the temperature value doesn't lie within the range), it guarantees that no \textit{false-negatives} are created (i.e., no records are filtered out that contain temperature measures). 
The \textit{false-positives}, on the contrary, are not affecting the end result, as the CPU parser will accurately filter the remaining records. 
Accordingly, \textit{false-positives} only reduce the achievable speedup gained from the filter selectivity.
However, the restriction to string matching poses an issue. 
Many  applications,  especially  from the  IoT  domain,  are  gaining  their  selectivity  from  filtering numbers, timestamps, or number ranges as exemplified in \Cref{lst:running_exp_q}. 

In this paper, we investigate raw filtering on FPGAs. 
Particularly, we introduce FPGA-based concepts for string matching and filtering number ranges in byte streams. 
Such raw filters can be combined to create more complex and ideally more selective filters, 
e.g., a conjunctive (AND) combination of string matching ($s("temperature")$) and a value range filter ($0.7 \le v \le 35.1$) for the query in \Cref{lst:running_exp_q}. 
However, filters built by conjunction and disjunction of filter predicates might not always result in improved selectivity. 
For example, the record in \Cref{lst:running_exp_rec} contains the "temperature" string and numbers ("$12$", "$20$") which lie in the given value range. 
However, the temperature value itself ("$35.2$") exceeds it. 
Thus, the described raw filter would generate a \textit{false-positive}. 
We therefore additionally present a technique which is able to extract structural information while scanning the byte stream. 
By using this technique, filter predicates can be combined such that their results are only combined if found in the correct structural context.

The \emph{contributions} of this paper can be summarized as follows:

\begin{enumerate}
    \item FPGA-based raw filtering concepts: \textit{primitives optimized for FPGA-based \glspl{rf}: (a) An approximate string matcher.
    (b) An approximate number range matcher.
    Extracting structural information and building raw filters by conjunctive and disjunctive combination of filter primitives depending on structural context.
    }

    \item Raw filter evaluation: \textit{query-specifc raw filter configurations: tradeoffs between resource requirements and false-positive rates}
    \item FPGA hardware architecture and implementation for raw filtering: 
    \textit{
    Filtering data streams at 10 GBit/s (line rate), i.e., no negative effect on performance of overall system.
    Tailored to filter the data stream directly between the data source (e.g., Ethernet, SSD) and CPU (SmartNIC, etc.). }
\end{enumerate}




The remaining paper is organized as follows: 
In \Cref{sec:Filter Architecture}, concepts and techniques for different raw filtering primitives are introduced and their composition into a complex raw filter is discussed.
In \Cref{sec:Raw Filter Evaluation}, the proposed filters are evaluated with the RiotBench Benchmark. 
Finally, the paper finishes in \Cref{sec:Conclusion and future work} with a conclusion and an outlook for future work.

\section{Related work}
\label{sec:Related work}

There has been a lot of research to accelerate parsing recently \cite{mison,simd-json}. Those implementations rely on SIMD-instructions and can scan documents at a speed of over 2 GB/s.
One of these Parsers is Mison \cite{mison} presented by \citeauthor{mison}. Mison uses SIMD-instructions to build a structural index of a record. Queried Fields are then parsed on demand by speculatively jumping to the beforehand indexed attribute positions.

Sparser \cite{sparser} was the first to approximately pre-filter records using SIMD-based string comparisons on raw data, to reduce the parsing workload itself.
Sparser's \glspl{rf} are based on two filtering primitives. The main primitive is a substring search, that inspects the input byte stream for 2-, 4- and 8-byte long substrings. The second primitive is used for key-value search and can filter co-occurrences of substrings.
One major drawback of Sparser is the limitation to filter only based on string comparisons. Many applications, however, especially those in the IoT domain, are gaining their selectivity from filtering numbers or timestamps.
In other scenarios, it's necessary to have at least a small degree of structural awareness, which is except for the key-value search not possible with Sparser.
The concept of raw filtering has been adopted for programmable switches, too \cite{FastStringSearchingOnPISA}. 
Here, strings are searched by several small cascaded DFAs. Nevertheless, programmable switches have limited resources and packets need to be recirculated through the switch, to filter complex expressions, therefore implying bandwidth limitations.

There have been several stream processing Query Compilers for FPGAs \cite{streams_on_wires,glacier,multi_query_stream_processing_on_FPGAs}. But these compilers solely work on binary data, not considering that the input data might have to be parsed before processing.
%
The Fleet \cite{Fleet} framework can push arbitrary Stream operations, like JSON Parsing, onto the FPGA by introducing a new DSL. However, their presented PUs can't process one character per cycle and must be replicated extensively to achieve a high throughput, hence requiring a lot of resources. 
%
ACCORDA \cite{ACCORDA} tries to improve the processing of raw unstructured data with dedicated Hardware Accelerators. The authors show that their unstructured data processor can parse and filter JSON data for all common predicates, but is in return again very resource intensive. The presented FPGA prototype requires 295K LUTs.
%
Moreover, regular expression matching has already been investigated for FPGAs heavily. \citet{woods_regex} suggest regular expression matching with an FPGA located between NIC and CPU, for complex event detection in data streams. As regular expressions can get arbitrarily complex and FPGA resources are limited, \citet{becher_regex} propose a optimistic regex evaluation. Compared to our raw filters, this approach enables similar tradeoffs between resource consumption and the accuracy of the filter. Nevertheless, we only used regular expressions for building our value range filter, as the resource consumption can be further reduced by using the other presented primitives.

\section{Raw Filtering Primitives \& Composition}
\label{sec:Filter Architecture}

In this section, we present FPGA filter primitives for string matching and for detecting values and value ranges. 
We furthermore propose a technique to extract some information of the JSON structure from the byte stream and describe how to compose these primitives to build query-specific selective raw filters. 

\subsection{String Search}


\begin{table*}\centering
    \vspace{-0.1cm}
    \caption{Comparison of the three string matching techniques for the SmartCity dataset.}\label{tab:sys_blocklen_comparison}    
    \vspace{-0.2cm}
    \csvreader[%
    tabular=|r |*{2}{| c c|}*{4}{| c c }|,
    respect underscore=true,
    table head=\hline \multirow{2}{*}{{search string}} & \multicolumn{2}{c||}{(i) DFA} & \multicolumn{2}{c||}{(ii) $N$-byte string} & \multicolumn{8}{c|}{(iii) $B$-byte substrings} \\
     & \multicolumn{2}{c||}{}& \multicolumn{2}{c||}{} & \multicolumn{2}{c}{$B=1$} & \multicolumn{2}{c}{$B=2$} & \multicolumn{2}{c}{$B=3$} & \multicolumn{2}{c|}{$B=4$}  \\
     & FPR & LUTs & FPR & LUTs & FPR & LUTs & FPR & LUTs & FPR & LUTs & FPR & LUTs\\\hline\hline,
    late after line=\\\hline
    ]%
    {data/sys_strings.dat}{}%
    {\csvlinetotablerow}%
\end{table*}
\begin{table*}\centering
    \vspace{-0.1cm}
    \caption{Comparison of the three string matching techniques for the Taxi dataset.}\label{tab:taxi_blocklen_comparison}
    \vspace{-0.2cm}
    \csvreader[%
    tabular=|r |*{2}{| c c|}*{4}{| c c }|,
    respect underscore=true,
    table head=\hline \multirow{2}{*}{{search string}} & \multicolumn{2}{c||}{(i) DFA} & \multicolumn{2}{c||}{(ii) $N$-byte string} & \multicolumn{8}{c|}{(iii) $B$-byte substrings} \\
     & \multicolumn{2}{c||}{}& \multicolumn{2}{c||}{} & \multicolumn{2}{c}{$B=1$} & \multicolumn{2}{c}{$B=2$} & \multicolumn{2}{c}{$B=3$} & \multicolumn{2}{c|}{$B=4$}  \\
     & FPR & LUTs & FPR & LUTs & FPR & LUTs & FPR & LUTs & FPR & LUTs & FPR & LUTs\\\hline\hline,
    late after line=\\\hline
    ]%
    {data/taxi_strings.dat}{}%
    {\csvlinetotablerow}%
\end{table*}
\begin{table*}\centering
    \vspace{-0.1cm}
    \caption{Comparison of the three string matching techniques for the Twitter dataset.}\label{tab:twitter_blocklen_comparison}
    \vspace{-0.2cm}
    \csvreader[%
    tabular=|r |*{2}{| c c|}*{4}{| c c }|,
    respect underscore=true,
    table head=\hline \multirow{2}{*}{{search string}} & \multicolumn{2}{c||}{(i) DFA} & \multicolumn{2}{c||}{(ii) $N$-byte string} & \multicolumn{8}{c|}{(iii) $B$-byte substrings} \\
     & \multicolumn{2}{c||}{}& \multicolumn{2}{c||}{} & \multicolumn{2}{c}{$B=1$} & \multicolumn{2}{c}{$B=2$} & \multicolumn{2}{c}{$B=3$} & \multicolumn{2}{c|}{$B=4$}  \\
     & FPR & LUTs & FPR & LUTs & FPR & LUTs & FPR & LUTs & FPR & LUTs & FPR & LUTs\\\hline\hline,
    late after line=\\\hline
    ]%
    {data/twitter_strings.dat}{}%
    {\csvlinetotablerow}%
    \vspace{-0.4cm}
\end{table*}

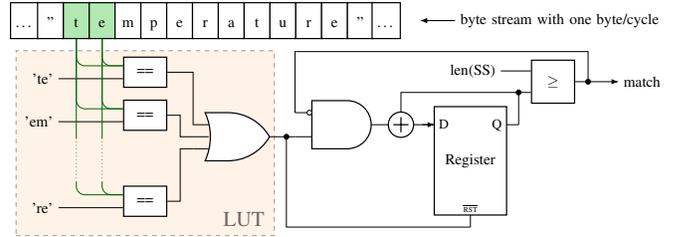
\begin{figure}

    \resizebox{\linewidth}{!}{%
        \begin{circuitikz}[european,tight background]

\ctikzset{
    logic ports=ieee,
}

\tikzstyle{background}=[fill=orange!10,draw=black!40, dashed]

\pgfdeclarelayer{background}
\pgfdeclarelayer{foreground}
\pgfsetlayers{background,main,foreground}

\tikzstyle{sline}=[draw, black!60!green,rounded corners=.2cm]

\foreach \i/\c/\y in {0/white/\ldots,1/white/",2/black!30!green!30/t, 3/black!30!green!30/e, 4/white/m, 5/white/p, 6/white/e, 7/white/r, 8/white/a, 9/white/t, 10/white/u, 11/white/r, 12/white/e, 13/white/", 14/white/\ldots}
    \node[draw,thick,fill=\c,
    minimum width=0.6cm,
    minimum height=0.6cm,
    align=left,
    text height=0.45cm, 
    text depth=0.15cm]
    (t\i) at (0.6*\i,0.2) {\y};

\coordinate (a) at (0, 0);
\coordinate (char_s) at (0.5, 0);
\coordinate (c) at (5, 0);

\draw [{Latex[length=1.5mm]}-]  ($(t14) + (0.8,0)$) -- ++(0.8,0) node[right](arr){byte stream with one byte/cycle};

\node [draw,thick,fill=white, anchor=center, minimum size=0.7cm, minimum width=1cm](eq0) at (2.8,-1) {==};
\coordinate (eq0in1) at ($(eq0.west) + (0,0.15)$);
\coordinate (eq0in2) at ($(eq0.west) - (0,0.15)$);
\coordinate (tmp) at ($(t2 |- eq0in1)$);
\draw[sline] (t2) -- ($(t2 |- eq0in1)$) -- (eq0in1);
\draw[sline] (t3) -- ($(t3 |- eq0in1)$) -- (eq0in1);
\draw (eq0in2) -- ++(-1.5,0)node[left](In1){'te'};

\node [draw,thick,fill=white, anchor=center, minimum size=0.7cm, minimum width=1cm](eq1) at ($(eq0)-(0,1)$) {==};
\coordinate (eq1in1) at ($(eq1.west) + (0,0.15)$);
\coordinate (eq1in2) at ($(eq1.west) - (0,0.15)$);
\coordinate (tmp) at ($(t2 |- eq1in1)$);
\draw[sline] (t2) -- ($(t2 |- eq1in1)$) -- (eq1in1);
\draw[sline] (t3) -- ($(t3 |- eq1in1)$) -- (eq1in1);
\draw (eq1in2) -- ++(-1.5,0)node[left](In1){'em'};

\node [draw,thick,fill=white, anchor=center, minimum size=0.7cm, minimum width=1cm](eq2) at ($(eq1)-(0,2)$) {==};
\coordinate (eq2in1) at ($(eq2.west) + (0,0.15)$);
\coordinate (eq2in2) at ($(eq2.west) - (0,0.15)$);
\coordinate (to2) at ($(t2 |- eq2in1) + (0,0.3)$);
\coordinate (to3) at ($(t3 |- eq2in1) + (0,0.3)$);
\draw[sline, dotted] (to2) -- ($(to2)+(0,1)$);
\draw[sline]         (t2) -- ($(to2)+(0,1)$);
\draw[sline, dotted] (to3) -- ($(to3)+(0,1)$);
\draw[sline]         (t3) -- ($(to3)+(0,1)$);
\draw[sline] (to2) -- ($(t2 |- eq2in1)$) -- (eq2in1);
\draw[sline] (to3) -- ($(t3 |- eq2in1)$) -- (eq2in1);
\draw (eq2in2) -- ++(-1.5,0)node[left](In1){'re'};

\node[or port, fill=white] (or1) at (5,-2.5){};
\draw (eq0.east) -- ($(eq0.east)+(0.6,0)$) |- (or1.in 1);
\draw (eq1.east) -- ($(eq1.east)+(0.3,0)$) |- ($(or1.west)!.33!(or1.center)$);
\draw (eq2.east) -- ($(eq2.east)+(0.3,0)$) |- (or1.in 2);

\node[and port, anchor=in 2] (and1) at ($(or1.out)+(0.2,0)$){};
\draw (or1.out) |- (and1.in 2);
\node at (and1.bin 1) [ocirc, left]{} ;

\node[adder, scale=0.6] (add1) at ($(and1.out)+(0.3,0)$){};
\draw (and1.out) |- (add1.west);

\tikzset{flipflop myD/.style={flipflop,
flipflop def={t1=D, t6=Q, td={\ctikztextnot{RST}}}}
};
\node[flipflop myD, anchor=pin 1](ff) at ($(add1.east)+(0.2,0)$){Register};
\draw[-{Latex[length=1.5mm]}] (add1.out) -- (ff.bpin 1);

\draw (ff.down) -- (or1.out|-ff.down) to[short,-*] (or1.out) ;

\node [draw,thick,fill=white, anchor=center, minimum size=1cm, minimum width=1cm](eq3) at ($(ff.pin 6)+(0.8,1)$) {$\geq$};
\coordinate (eq3in1) at ($(eq3.west) + (0,0.25)$);
\coordinate (eq3in2) at ($(eq3.west) - (0,0.25)$);
\draw (eq3in1) -- ++(-0.7,0)node[left](In1){len(SS)};
\draw[-{Latex[length=1.5mm]}] (eq3.east) -- ++(1,0)node[right](In1){match};

\coordinate (tmp) at ($(ff.pin 6 |- eq3in2)$);
\draw (ff.pin 6) -- (tmp) -- (eq3in2);
\draw (add1.north) -- (add1.north|-tmp) to[short,-*] (tmp);

\coordinate (tmp) at ($(eq3.east)+(0.3,0.7)$);
\draw (and1.in 1) -- ($(and1.in 1 |- tmp)$) -- (tmp) to[short,-*] ($(tmp |- eq3.east)$); %

\begin{pgfonlayer}{background}
    \path (eq0)+(-3,0.5) node (a) {};
    \path ($(or1 |- eq2)$)+(+0.8,-0.8) node (b) {};
    \coordinate (box0) at ($(a|-b)$);
    \path[background]
        (a) rectangle (b);        
\end{pgfonlayer}
\node[anchor=south east, color=black!60] (lut) at ($(b)+(-0.1,0.1)$){\Large{}LUT};

\end{circuitikz}
    }
    \caption{RTL architecture for a "temperature" string search using a block length of $B=2$}
    \label{string_matcher_architecture}
    \vspace{-0.4cm}

\end{figure}

The original idea of raw filtering is to find a string anywhere in a given byte stream regardless of any structure. 
In the following, we investigate three techniques to implement string matching on FPGAs. 
Let the search string have a size of $N$ Bytes. 
(i) The string can be matched with a state machine accepting the search string with $N$ states, which transitions with one character every cycle.
(ii) Another option would be to buffer the last $N$ Bytes of the byte stream and compare them to the full $N$-Byte search string every cycle.
While these solutions exactly match a given search string, we want to introduce a more resource-saving approximate option that allows rare \textit{false-positives} to occur. 
(iii) Instead of buffering the full length of the search string, we only buffer the last \gls{b} Bytes of the input byte stream. 
This buffer is compared to all possible substrings of a size of \gls{b} Bytes of the search string. 
The results of these comparators are then or-reduced and fed into a counter, which is incremented with each match and reset when no comparator matches. 
The filter emits a '1' if the counter value is equal to $N-B+1$. 
As an example, \Cref{tab:b_substring_sets} shows all substrings of the search string "temperature" for different block lengths \gls{b}.
\Cref{string_matcher_architecture} presents a schematic of our concept for search string "temperature" and block length $B=2$. 
In the remainder of this paper, searching a string $str$ with block length $B$ is denoted as $s_{B}(str)$.


\begin{table}[t]
    \vspace{0.1cm}
    \caption{Substrings of the "temperature" search string for different block lengths $B$. Duplicates are indicated with ().}
    \label{tab:b_substring_sets}
    \resizebox{\linewidth}{!}{%
        \begin{tabular}{l l}
            \toprule
            $B$      & sub-strings                                                   \\ \midrule
            $1$      & 't', 'e', 'm', 'p', ('e',) 'r', 'a', ('t',) 'u', ('r', 'e')   \\
            $2$      & 'te', 'em', 'mp', 'pe', 'er', 'ra', 'at', 'tu', 'ur', 're'    \\
            $3$      & 'tem', 'emp', 'mpe', 'per', 'era', 'rat', 'atu', 'tur', 'ure' \\
            \vdots & \vdots                                                        \\
            $n$      & 'temperature'                                                 \\ \bottomrule
        \end{tabular}
    }
    \vspace{-0.5cm}
\end{table}


In the following, these three techniques will be evaluated in terms of \gls{fpr} and resource costs (LUTs).
For this purpose, the strings used in RiotBench \cite{riotbench} are examined with their associated datasets.
As these measurement lists aren't containing many strings, which might cause \textit{false-positives}, we additionally evaluated a more diverse Twitter dataset \cite{Go2009SentimentCU}. 
The results can be seen in \Cref{tab:sys_blocklen_comparison,tab:taxi_blocklen_comparison,tab:twitter_blocklen_comparison}. 
While for the exact solutions (i, ii) the required LUTs increase rapidly with the length of the string, only slightly more resources are required for longer strings for the substring search (iii). 
This is due to the or-reduction of the many comparisons, which ensures that the entire logic can be combined in one LUT. 
This large LUT can be mapped effectively to hardware primitives by the synthesis tool.
In the comparison of the exact methods (i, ii), the full-length comparison (ii) turns out to be advantageous for shorter strings, since no states have to be encoded. 
However, as the length of the strings increases, the amount of logic required to process several chars in parallel increases significantly. 
In contrast, the DFA-based solution (i) only requires one char to be processed per cycle, while the number of bits required to encode the states increases only approximately logarithmically.
With respect to the required LUTs, the presented subset matcher with $B=1$ turns out to be superior in all cases. 
But especially for longer strings, where even for $B=1$ an FPR of $0$ is achieved. 
For shorter strings, however, an FPR of 0 is only achieved for $B=2$, whereby the full-length comparisons in these cases partially require fewer LUTs. 
In rare cases, with a $B=1$, even longer strings can be completely confused with other regularly occurring strings, as can be seen for the string $s_{1}("tolls\_amount")$, which contains the same letters as the string "total\_amount". 
Here, however, a favorable solution can be found with a $B=2$.
All in all, there are three interesting variants for our \glspl{rf}. 
For long strings, an FPR of 0 can usually be achieved for $B=1$ with a minimum of resources. 
For all other cases $B=2$ is usually advantageous, but in some cases a full length comparison is also preferable. 
Accordingly, the search space for following evaluations of the RiotBench concentrates on $ B \in \{1, 2, N\}$, where $B=N$ corresponds to string matching option (ii).

\subsection{Number Range Filtering}

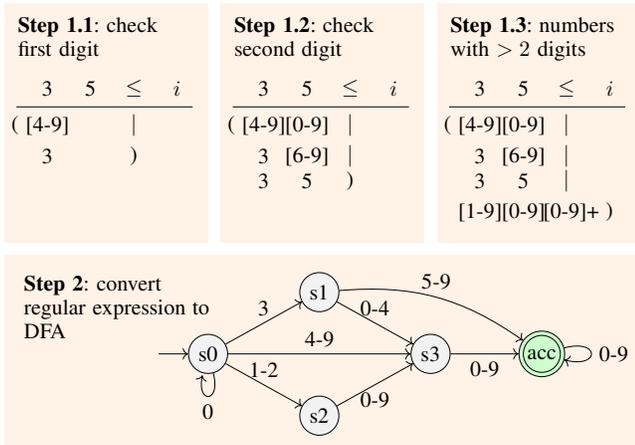
\begin{figure}
    \centering
    \resizebox{\linewidth}{!}{%
        \begin{tikzpicture}
    \pgfdeclarelayer{background}
    \pgfdeclarelayer{foreground}
    \pgfsetlayers{background,main,foreground}
    
    \usetikzlibrary{calc}
    \tikzstyle{background}=[fill=orange!10]
        
    \def\blockdist{1.4}
    
    \def\digitdist{0.7}
    \def\eqdist{0.6}
    
    \def\linedist{0.3}
    \def\lineover{0.5}
    
    \def\rowdist{0.5}
    \def\chardist{0.4}
    \def\bracedist{0.05}
    
    \def\boxheight{-2.5}
    \def\nodedist{0.8}

    \node (d0) at (0,0) {3} ;
    \path (d0)+(+\digitdist,0) node (d1) {5};
    \path (d1)+(+\digitdist,0) node (eq) {$\leq$};
    \path (eq)+(+\digitdist,0) node (x) {$i$};
    
    \path (d0)+(0,-\eqdist) node (t0) {[4-9]};
    \path (t0)+(0,-\rowdist) node (t1) {3};
    
    \path [draw] ($(d0) + (-\lineover,-\linedist)$) -- ($(x|-d0) + (0.5*\lineover,-\linedist)$);
    
    \path (d0)+(-0.6,\nodedist) node [anchor=west, text width=2.8cm] (label) {\textbf{Step 1.1}: check first digit};
    
    \path (t0.west)+(-\bracedist,0) node (b1) {(};
    \path ($(eq|-t0)$) node (a) {$|$};
    \path ($(eq|-t1)$) node (a) {)};
        
    \begin{pgfonlayer}{background}
        \path ($(label.west|-label.north)$)+(-0.1,0.1) node (a) {};
        \path (x)+(+0.5,\boxheight) node (b) {};
        \coordinate (box0) at ($(a|-b)$);
        \path[background]
            (a) rectangle (b);
            
    \end{pgfonlayer}
    
    \path (x)+(+\blockdist,0) node (d0) {3};
    \path (d0)+(+\digitdist,0) node (d1) {5};
    \path (d1)+(+\digitdist,0) node (eq) {$\leq$};
    \path (eq)+(+\digitdist,0) node (x) {$i$};
    
    \path (d0)+(0,-\eqdist) node (t0) {[4-9]};
    \path (t0)+(0,-\rowdist) node (t1) {3};
    
    \path (d1)+(0,-\eqdist) node (t10) {[0-9]};
    \path (t10)+(0,-\rowdist) node (t11) {[6-9]};
    
    \path (t1)+(0,-\chardist) node (k1) {3};
    \path (t11)+(0,-\chardist) node (t12) {5};
    
    \path (t0.west)+(-\bracedist,0) node (b1) {(};
    
    \path ($(eq|-t0)$) node (a) {$|$};
    \path ($(eq|-t1)$) node (a) {$|$};
    \path ($(eq|-t12)$) node (a) {)};
    
    \path [draw] ($(d0) + (-\lineover,-\linedist)$) -- ($(x|-d0) + (0.5*\lineover,-\linedist)$);
    
    \path (d0)+(-0.6,\nodedist) node [anchor=west, text width=2.8cm] (label) {\textbf{Step 1.2}: check second digit};
        
    \begin{pgfonlayer}{background}
        \path ($(label.west|-label.north)$)+(-0.1,0.1) node (a) {};
        \path (x)+(+0.5,\boxheight) node (b) {};
        \path[background]
            (a) rectangle (b);
            
    \end{pgfonlayer}
    
    \path (x)+(+\blockdist,0) node (d0) {3};
    \path (d0)+(+\digitdist,0) node (d1) {5};
    \path (d1)+(+\digitdist,0) node (eq) {$\leq$};
    \path (eq)+(+\digitdist,0) node (x) {$i$};
    
    \path (d0)+(0,-\eqdist) node (t0) {[4-9]};
    \path (t0)+(0,-\rowdist) node (t1) {3};
    
    \path (d1)+(0,-\eqdist) node (t10) {[0-9]};
    \path (t10)+(0,-\rowdist) node (t11) {[6-9]};
    
    \path (t1)+(0,-\chardist) node (t3) {3};
    \path (t11)+(0,-\chardist) node (t12) {5};
    
    \path (t3)+(0,-\rowdist) node (p0) {[1-9]};
    \path ($(t10|-p0)$) node (p1) {[0-9]};
    \path (p1)+(+1.1*\digitdist,0) node (p2) {[0-9]+};
    
    \path (t0.west)+(-\bracedist,0) node (b1) {(};
    \path (p2.east)+(+\bracedist,0) node (b1) {)};
    
    \path ($(eq|-t0)$) node (a) {$|$};
    \path ($(eq|-t1)$) node (a) {$|$};
    \path ($(eq|-t12)$) node (a) {$|$};
    
    \path [draw] ($(d0) + (-\lineover,-\linedist)$) -- ($(x|-d0) + (0.5*\lineover,-\linedist)$);
    
    \path (d0)+(-0.6,\nodedist) node [anchor=west, text width=2.8cm] (label) {\textbf{Step 1.3}: numbers with $> 2$ digits};
        
    \begin{pgfonlayer}{background}
        \path ($(label.west|-label.north)$)+(-0.1,0.1) node (a) {};
        \path (x)+(+0.5,\boxheight) node (b) {};
        \coordinate (box1) at (b);
        \path[background]
            (a) rectangle (b);
            
    \end{pgfonlayer}
    
    \coordinate (a) at ($(box0) + (0,-0.2)$);
    \coordinate (b) at ($(box1) + (0,-4.2)$);
    \path (a)+(0.2,-0.2) node [anchor=north west, text width=3cm] (label) {\textbf{Step 2}: convert regular expression to DFA};
        
    \tikzstyle{vertex}=[circle,fill=black!5,minimum size=18pt,draw,inner sep=0pt]
    \tikzstyle{rp}=[->,color=red!60]
        
    \tikzstyle{accepting}=[path picture={%
      \draw let 
        \p1 = (path picture bounding box.east),
        \p2 = (path picture bounding box.center)
        in
          (\p2) circle (\x1 - \x2 - 2pt);
      }]
    
    \def\vdist{1.8}
    \def\hvdist{1}
    
    \coordinate (dfa) at ($(a) + (3.3,-1.6)$);
    \node[vertex] (s0) at (dfa){s0};
    \node[vertex] (s1) at ($(s0) + (\vdist,\hvdist)$){s1};
    \node[vertex] (s2) at ($(s0) + (\vdist,-\hvdist)$){s2};
    \node[vertex] (s3) at ($(s1|-s0) + (\vdist,0)$){s3};
    \node[vertex,accepting,minimum size=21pt, fill=green!20] (acc) at ($(s3) + (\vdist,0)$){acc};
    
  
    \draw[<-] (s0) -- ++(-0.8,0);
    
    \draw [->]  (s0) edge [loop below] node[below] {0} (s0);
    \draw [->]  (s0) to node[above] {3} (s1);
    \draw [->]  (s0) to node[above] {1-2} (s2);
    \draw [->]  (s0) to node[above] {4-9} (s3);
    \draw [->]  (s2) to node[below] {0-9} (s3);
    \draw [->]  (s1) to node[above] {0-4} (s3);
    \draw [->]  (s3) to node[below] {0-9} (acc);
    \draw [->, bend angle=20, bend left]  (s1) to node[above] {5-9} (acc);
    
    \draw [->, loop right]  (acc) to node[right] {0-9} (acc);
    
    \begin{pgfonlayer}{background}
        \coordinate (b) at ($(b|-s2) + (0,-0.5)$);
        \path[background]
            (a) rectangle (b);
            
    \end{pgfonlayer}
\end{tikzpicture}
    }
    \caption{Number filter build process example for $i \geq 35$. Step 1: a regular expression is derived from the value comparison. Step 2: the regex is transformed into a DFA. The green vertex indicates an accepting state.}
    \label{fig:value_range_regex}
    \vspace{-0.3cm}
\end{figure}

For scanning a byte stream to contain specific integer, float, or time stamp values or even value ranges, we propose to filter the stream with \glspl{dfa}. 
For this, a regular expression for the value (range) is first derived and then converted into a \gls{dfa} to be synthesized for the FPGA. 
\Cref{fig:value_range_regex} illustrates the process for detecting integers greater or equal to $35$ ($i \geq 35$) in a byte stream. 
The same approach is used for deriving a regex for upper bounds (e.g., $i \leq 35$). 
The comparison against a range of values, i.e. with an upper and a lower bound, can still be performed with only one automaton, which can later be optimized better than two separate automata and thus requires fewer resources overall. 

Building these \glspl{dfa} for floating-point numbers is done similarly. After the decimal point, we just continue to check the given bounds for the decimal places. 
Nevertheless, there is one major difference between floats and integers which can't be mapped into \glspl{dfa} easily. 
The JSON specification allows an exponent format (e.g., 2.1e3), which is virtually impossible to match via state machines. 
That's because of the almost unlimited possibilities to format the same number (e.g., 1e+1, 10, 100e-1, \ldots). 
As a remedy, we accept any number, including at least one digit followed by an 'e' or an 'E'. 
In such a case, the raw filter primitive might create a \textit{false-positive}, but at least no \textit{false-negative}. 

In step 2 the regular expression is converted into a \gls{dfa} and minimized.
Methods to achieve this are already well known and won't be explained here. 
The \gls{dfa} is evaluated every time a non-numeric (including '+', '-', '.', 'e') character is seen, as it has to mark the end of the number. 
At this point, the value filter omits a '1' if the last state was an accepted state, or a '0' if the last state wasn't accepted. 
The \gls{dfa} is then reseted and starts over from state s0.
The shown method is not only valid for numerical filters, but can also be used for date formats or any other filter which can be represented using regular expressions.
A number matcher for an integer $i$ or a float $f$ with a lower bound $\ell$ and an upper bound $u$ is denoted in the following as $\ell \le i \le u$ or $\ell \le f \le u$, respectively.



\subsection{Structural awareness}

In many scenarios, applications require some degree of \textit{structural awareness}. However, this doesn't mean that we have to parse the full structure of a given record. Instead, we propose to observe only those elements which help us the most with reducing the number of \textit{false-positives}. 


As already shown in the introductory example, it can be advantageous to observe object structures, especially for filtering IoT data, since sensor values are commonly stored together with metadata in an object. 
This is also evident for the \gls{senml} standard used in running example (\Cref{lst:running_exp_rec}), where a JSON record is defined over an array of sensor measurements. 
Since such sensor-measurement objects usually contain the same attributes for all objects, key-value relationships are hardly relevant for filtering (see~\Cref{lst:running_exp_rec} where the keys "v", "u", and "n" appear in all objects of the record). 
For more conventional JSON data, such as statistics gathered from web service APIs (e.g., Twitter\footnote{https://developer.twitter.com/en/docs/twitter-api}), key-value relationships are much more important, since the keys are usually only represented once in the entire record and can thus be used as a unique identifier.



This sensitivity for nesting levels is achieved by incrementing a counter with every '[','\{'and decrementing it with every '\}',']'.
As this counter has to be consistent at any time, it's necessary to detect if a bracket is part of a string and should therefore not be used to alter the nesting level. 
Detecting strings, however, requires checking if a quote " is escaped by a '\textbackslash' character. 
And \textbackslash can again be escaped by \textbackslash\textbackslash. 
This information can then be used to build a string mask and consequently to correctly determine the nesting levels.

Detecting key-value pairs can be done using the same method. 
Instead of tracking the nesting levels we just need to check that the key \gls{rf} and the value \gls{rf} both appear before the same unescaped comma.
Two \glspl{rf} (RF1, RF2) appearing on the same nesting level are denoted as \{RF1 \& RF2\} in the following.

\subsection{Design Flow}
\label{sec:Design Flow}


The design flow for generating an optimized raw filter for a given query can be broken down into four steps. 
i) Extract search strings and value ranges from the query.
ii) Select corresponding primitives and their parameters that should be tested (in our case: values for block length $B$).
iii) Determine possible combinations that constitute the design space:  
(a) Primitives that appear in the same context could be combined via structural-aware filtering or without. 
(b) Primitives in \textit{and}-clauses can be omitted to reduce the raw filter size as long as one primitive remains.
For \textit{or}-clauses, however, all input expressions must be considered, to prohibit \textit{false-negatives}.
Finally, iv) the design space is explored to determine design points that are optimal, e.g., with respect to \gls{fpr} and resource requirements.  
\Cref{sec:Raw Filter Evaluation} exemplifies the design flow for an IoT benchmark.


\section{Raw Filter Evaluation}
\label{sec:Raw Filter Evaluation}
In the following, we evaluate the most interesting design points for three different queries and present an example system that implements the best implementation for one of these queries.

\subsection{Analysis of the Design Space}

For our evaluations, we used three different queries operating on two different datasets, from the RiotBench. 
The queries QS0 \& QS1 run on a SmartCity dataset, composed of different sensor measurements (e.g., temperature, humidity). 
The third Query QT uses a Taxi dataset, where each record corresponds to one Taxi trip (containing attributes like trip distance, fare amount). 
The queries are shown in \Cref{tab:riotbench_queries}.
%
%
\begin{table}
    \centering
    \caption{Pareto points for QS0}
    \renewcommand\tabularxcolumn[1]{m{#1}}
    \begin{tabularx}{0.96\linewidth}{ X c c }\toprule
        \textbf{Raw-filter configuration} & \textbf{FPR} & \textbf{LUTs} \\\midrule\midrule
        $v(12 \leq i \leq 49)$ & 0.853 &    18 \\ \hline
\{ $s_{1}("airquality\_raw")$ \& $v(12 \leq i \leq 49)$ \} & 0.770 &    47 \\ \hline
\{ $s_{1}("humidity")$ \& $v(20.3 \leq f \leq 69.1)$ \} & 0.562 &    95 \\ \hline
\{ $s_{1}("humidity")$ \& $v(20.3 \leq f \leq 69.1)$ \} \newline \& \{ $s_{1}("airquality\_raw")$ \& $v(12 \leq i \leq 49)$ \} & 0.349 &   123 \\ \hline
\{ $s_{1}("temperature")$ \& $v(0.7 \leq f \leq 35.1)$ \} \newline \& \{ $s_{1}("humidity")$ \& $v(20.3 \leq f \leq 69.1)$ \} \newline \& $v(12 \leq i \leq 49)$ & 0.266 &   151 \\ \hline
\{ $s_{1}("temperature")$ \& $v(0.7 \leq f \leq 35.1)$ \} \newline \& \{ $s_{1}("humidity")$ \& $v(20.3 \leq f \leq 69.1)$ \} \newline \& \{ $s_{1}("airquality\_raw")$ \& $v(12 \leq i \leq 49)$ \} & 0.208 &   172 \\ \hline
\{ $s_{1}("humidity")$ \& $v(20.3 \leq f \leq 69.1)$ \} \newline \& \{ $s_{1}("dust")$ \& $v(83.36 \leq f \leq 3322.67)$ \} \newline \& $v(12 \leq i \leq 49)$ & 0.205 &   204 \\ \hline
\{ $s_{1}("temperature")$ \& $v(0.7 \leq f \leq 35.1)$ \} \newline \& \{ $s_{1}("humidity")$ \& $v(20.3 \leq f \leq 69.1)$ \} \newline \& \{ $s_{1}("light")$ \& $v(0 \leq i \leq 5153)$ \} \newline \& \{ $s_{1}("airquality\_raw")$ \& $v(12 \leq i \leq 49)$ \} & 0.197 &   211 \\ \hline
\{ $s_{1}("humidity")$ \& $v(20.3 \leq f \leq 69.1)$ \} \newline \& \{ $s_{1}("dust")$ \& $v(83.36 \leq f \leq 3322.67)$ \} \newline \& \{ $s_{1}("airquality\_raw")$ \& $v(12 \leq i \leq 49)$ \} & 0.144 &   220 \\ \hline
\{ $s_{1}("humidity")$ \& $v(20.3 \leq f \leq 69.1)$ \} \newline \& \{ $s_{1}("light")$ \& $v(0 \leq i \leq 5153)$ \} \newline \& \{ $s_{1}("dust")$ \& $v(83.36 \leq f \leq 3322.67)$ \} \newline \& \{ $s_{1}("airquality\_raw")$ \& $v(12 \leq i \leq 49)$ \} & 0.130 &   255 \\ \hline
\{ $s_{1}("temperature")$ \& $v(0.7 \leq f \leq 35.1)$ \} \newline \& \{ $s_{1}("humidity")$ \& $v(20.3 \leq f \leq 69.1)$ \} \newline \& \{ $s_{1}("dust")$ \& $v(83.36 \leq f \leq 3322.67)$ \} \newline \& $v(12 \leq i \leq 49)$ & 0.064 &   262 \\ \hline
\{ $s_{1}("temperature")$ \& $v(0.7 \leq f \leq 35.1)$ \} \newline \& \{ $s_{1}("humidity")$ \& $v(20.3 \leq f \leq 69.1)$ \} \newline \& \{ $s_{1}("dust")$ \& $v(83.36 \leq f \leq 3322.67)$ \} \newline \& \{ $s_{1}("airquality\_raw")$ \& $v(12 \leq i \leq 49)$ \} & 0.011 &   274 \\ \hline
\{ $s_{1}("temperature")$ \& $v(0.7 \leq f \leq 35.1)$ \} \newline \& \{ $s_{1}("humidity")$ \& $v(20.3 \leq f \leq 69.1)$ \} \newline \& \{ $s_{1}("light")$ \& $v(0 \leq i \leq 5153)$ \} \newline \& \{ $s_{1}("dust")$ \& $v(83.36 \leq f \leq 3322.67)$ \} \newline \& \{ $s_{1}("airquality\_raw")$ \& $v(12 \leq i \leq 49)$ \} & 0.000 &   307 \\ \hline
    \end{tabularx}
    \label{tab:sys_qo_pareto_tab}
    \vspace{-0.3cm}
\end{table}
\begin{table}
    \centering
    \renewcommand\tabularxcolumn[1]{m{#1}}
    \caption{Pareto points for QS1}
    \begin{tabularx}{0.96\linewidth}{ X c c }\toprule
        \textbf{Raw-filter configuration} & \textbf{FPR} & \textbf{LUTs} \\\midrule\midrule
        $v(17 \leq i \leq 363)$ & 0.964 &    35 \\ \hline
$v(1345 \leq i \leq 26282)$ & 0.130 &    38 \\ \hline
\{ $s_{1}("light")$ \& $v(1345 \leq i \leq 26282)$ \} & 0.029 &    75 \\ \hline
\{ $s_{1}("light")$ \& $v(1345 \leq i \leq 26282)$ \} \newline \& \{ $s_{1}("airquality\_raw")$ \& $v(17 \leq i \leq 363)$ \} & 0.008 &   103 \\ \hline
\{ $s_{1}("light")$ \& $v(1345 \leq i \leq 26282)$ \} \newline \& \{ $s_{1}("dust")$ \& $v(186.61 \leq f \leq 5188.21)$ \} \newline \& \{ $s_{1}("airquality\_raw")$ \& $v(17 \leq i \leq 363)$ \} & 0.000 &   223 \\ \hline
    \end{tabularx}
    \label{tab:sys_q1_pareto_tab}
    \vspace{-0.3cm}
\end{table}
\begin{table}
    \centering
    \renewcommand\tabularxcolumn[1]{m{#1}}
    \caption{Pareto points for QT}
    \begin{tabularx}{0.96\linewidth}{ X c c }\toprule
        \textbf{Raw-filter configuration} & \textbf{FPR} & \textbf{LUTs} \\\midrule\midrule
        $v(2.5 \leq f \leq 18.0)$ & 1.000 &    37 \\ \hline
$v(140 \leq i \leq 3155)$ & 0.998 &    62 \\ \hline
\{ $s_{1}("tolls\_amount")$ \& $v(2.5 \leq f \leq 18.0)$ \} & 0.722 &    65 \\ \hline
\{ $s_{2}("tolls\_amount")$ \& $v(2.5 \leq f \leq 18.0)$ \} & 0.021 &    81 \\ \hline
\{ $s_{2}("tip\_amount")$ \& $v(0.65 \leq f \leq 38.55)$ \} \newline \& \{ $s_{2}("tolls\_amount")$ \& $v(2.5 \leq f \leq 18.0)$ \} & 0.000 &   159 \\ \hline
    \end{tabularx}
    \label{tab:taxi_pareto_tab}
    \vspace{-0.5cm}
\end{table}
\begin{table*}
    \caption{RiotBench queries as used in the evalution.}
    \label{tab:riotbench_queries}
    \vspace{-0.2cm}

    \lstset {columns=fixed,
        language=predicate,
        escapeinside={(*}{*)},
        aboveskip=-1em,
        belowskip=-0.5em
        }    
    
    \begin{tabular}{p{48.6pt} p{0.74\linewidth} l}
        \toprule
        \textbf{Query Name} & \textbf{Filter expression} & \textbf{Selectivity (\%)}\\ \midrule
        SmartCity 0\newline (\textbf{QS0}) & \begin{lstlisting}
(0.7 (*$\leq$*) "temperature" (*$\leq$*) 35.1) AND (20.3  (*$\leq$*) "humidity" (*$\leq$*) 69.1)
AND (0   (*$\leq$*) "light" (*$\leq$*) 5153)   AND (83.36 (*$\leq$*) "dust" (*$\leq$*) 3322.67)
AND (12  (*$\leq$*) "airquality_raw" (*$\leq$*) 49)\end{lstlisting} & \textbf{63.9}\\

SmartCity 1\newline (\textbf{QS1}) & \begin{lstlisting}
(-12.5 (*$\leq$*) "temperature" (*$\leq$*) 43.1) AND (10.7 (*$\leq$*) "humidity" (*$\leq$*) 95.2)
AND (1345 (*$\leq$*) "light" (*$\leq$*) 26282)   AND (186.61 (*$\leq$*) "dust" (*$\leq$*)  5188.21)
AND (17 (*$\leq$*) "airquality_raw" (*$\leq$*) 363)
\end{lstlisting} & \textbf{5.4}\\

Taxi (\textbf{QT}) & \begin{lstlisting}
(140  (*$\leq$*) "trip_time_in_secs" (*$\leq$*) 3155) AND (0.65 (*$\leq$*) "tip_amount" (*$\leq$*) 38.55) 
AND (6.00 (*$\leq$*) "fare_amount" (*$\leq$*) 201.00) AND (2.50 (*$\leq$*) "tolls_amount" (*$\leq$*) 18.00)
AND (1.37 (*$\leq$*) "trip_distance" (*$\leq$*) 29.86)
\end{lstlisting} & \textbf{5.7}\\
        \bottomrule
    \end{tabular}
    \vspace{-0.2cm}
\end{table*}
\begin{figure*}
    \centering
    \begin{subfigure}{.33\textwidth}
        \centering
        \begin{tikzpicture}[scale=.58]
    \begin{axis}[
      colorbar ,
      colorbar sampled,
      colorbar style={
          samples=6,
          ylabel=Num Attributes
          },  
      point meta min=0.5,
      point meta max=5.5,
      xlabel=FPR,
      ylabel=Total LUTs,
      ]
      \addplot[
      only marks,
      scatter,
      scatter src=explicit,
      opacity=0.2,
      mark=*,
      mark size=2.9pt
      ]
      table[meta=na]{data/sys_q0_attr.dat};
    \end{axis}
  \end{tikzpicture}
        \caption{Smart City Query QS0}
        \label{fig:resource_scatter_sys0}
    \end{subfigure}%
    \begin{subfigure}{.33\textwidth}
        \centering
        \begin{tikzpicture}[scale=.58]
    \begin{axis}[ 
        colorbar ,
        colorbar sampled,
        colorbar style={
            samples=6,
            ylabel=Num Attributes
            },  
        point meta min=0.5,
        point meta max=5.5,
        xlabel=FPR,
        ylabel=Total LUTs,
        ]
        \addplot[
        only marks,
        scatter,
        scatter src=explicit,
        opacity=0.2,
        mark=*,
        mark size=2.9pt
        ]
        table[meta=na]{data/sys_q1_attr.dat};
    \end{axis}
\end{tikzpicture}
        \caption{Smart City Query QS1}
        \label{fig:resource_scatter_sys1}
    \end{subfigure}%
    \begin{subfigure}{.33\textwidth}
        \centering
        \begin{tikzpicture}[scale=.58]
    \begin{axis}[
        colorbar ,
        colorbar sampled,
        colorbar style={
            samples=6,
            ylabel=Num Attributes
            },  
        point meta min=0.5,
        point meta max=5.5,
        xlabel=FPR,
        ylabel=Total LUTs,
        ]
        \addplot[
        only marks,
        scatter,
        scatter src=explicit,
        opacity=0.2,
        mark=*,
        mark size=2.9pt
        ]
        table[meta=na]{data/taxi_attr.dat};
    \end{axis}
\end{tikzpicture}
        \caption{Taxi Query QT}
        \label{fig:resource_scatter_taxi}
    \end{subfigure}
    \caption{Scatter plot showing total LUT count and FPR for 3 Queries used in the RiotBench. The color of the points indicates the number of attributes which are filtered.}
    \label{fig:resource_scatter}
    \vspace{-0.2cm}
\end{figure*}
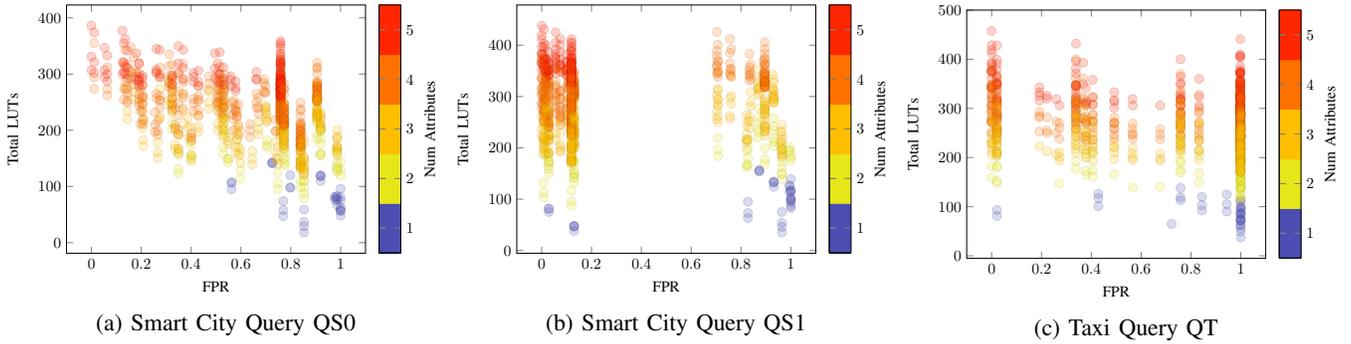
%
%
\label{sec:rf_analysis}


Following the design flow from \Cref{sec:Design Flow}, we extracted all primitives; i.e.,
five string matchers (each with the option to be implemented with $B \in \{1,2, N\}$) and five number range checks. 
Additionally, the information which RFs could utilize information about the structure is extracted. 
All possible configurations are then generated and individually evaluated with respect to resource requirements (LUTs on the FPGA) and \gls{fpr} over the complete data set. 
The tested design points are illustrated in \Cref{fig:resource_scatter}.
\Cref{tab:sys_qo_pareto_tab,tab:sys_q1_pareto_tab,tab:taxi_pareto_tab} provide the found pareto-optima.
For QS1 and QT, we can see that filtering for only 1-2 attributes already achieves a very low \gls{fpr}, as our queries are more selective over certain attributes. 
Since all attribute filters here are connected by an and, all but one filter may be omitted, as can be seen for the configurations obtained.
For IoT-Data it's quite common that some attributes highly correlate with each other.
E.g., in the taxi dataset the attributes \textit{trip\_time\_in\_secs} and \textit{fare\_amount} are highly dependent on the attribute \textit{trip\_distance}. 
Hence, it's sufficient to filter only one of these attributes. 
In other cases, we observed that value distributions of some attributes do not overlap with those of the other attributes. 
This can be seen for the \textit{"light"} attribute, where the light values are mostly $> 1000$, while the other attributes are mostly $< 1000$. 
Accordingly, it can be sufficient to only filter the value range without checking the attributes name. 
This can be seen for the second Pareto point in QS1 which already obtains a low \gls{fpr} without searching for the search string "light". 

If we look at the Pareto points independently of the underlying configurations, we notice that the FPRs and the LUTs cannot be mapped linearly to each other.
Especially at the ends of the tables, strong deviations can be seen. 
For example, for QS1, a minimally higher FPR of 0.008 compared to 0.0 requires less than half the resources (103 instead of 223 LUTs).
Accordingly, it may be worthwhile to allow a low FPR to save resources.



\subsection{System Architecture}
\label{sec:Experiments Results}

\begin{figure}
    \vspace{-0.2cm}
    \centering
\begin{tikzpicture}[
        tight background,
        >=stealth,
        database/.style={
            cylinder,
            shape border rotate=90,
            aspect=0.25,
            fill=ssdcolor,
        },
        operator/.style={
            rectangle,
            rounded corners,
            fill=ssdcolor!30,
            draw,
            minimum size=1em,
        },
        line width=.05em,
        scale=0.8,
        every node/.style={scale=0.8}
]
\pgfdeclarelayer{background}
\pgfdeclarelayer{foreground}
\pgfsetlayers{background,main,foreground}

\usetikzlibrary{arrows.meta}
\def\arroW{3mm}
\def\arroL{1.5mm}
\def\lineW{1.5mm}
\tikzstyle{big_arrow}=[draw=black!50,line width=\lineW, -{Triangle[length=\arroL,width=\arroW]}]
\tikzstyle{big_arrow_d}=[draw=black!50,line width=\lineW, {Triangle[length=\arroL,width=\arroW]}-{Triangle[length=\arroL,width=\arroW]}]

\tikzstyle{mid_arrow}=[draw=black!50,line width=0.3*\lineW, -{Triangle[length=0.5*\arroL,width=0.6*\arroW]}]
\def\blockdist{1em}

\def\blockHeight{6.75em}

\node [rectangle,draw,dashed,fill=staticcolor,minimum width=4em,minimum height=\blockHeight,text width=3.5em,align=center] at (0,0) (net) {Network\\interface}; 


\node [rectangle,draw,fill=prcolor,minimum width=6em,minimum height=\blockHeight,right=\blockdist of net] (pr3) {};

\node [rectangle,draw,fill=corecolor,minimum width=2.5em,minimum height=1.15em,yshift=-0.1em,below left=0.1em of pr3.north] (rf0) {\scriptsize RF};
\node [rectangle,draw,fill=corecolor,minimum width=2.5em,minimum height=1.15em,yshift=-0.1em,below right=0.1em of pr3.north] (rf1) {\scriptsize RF};

\node [rectangle,draw,fill=corecolor,minimum width=2.5em,minimum height=1.15em,below=0.1em of rf0] (rf2) {\scriptsize RF};
\node [rectangle,draw,fill=corecolor,minimum width=2.5em,minimum height=1.15em,below=0.1em of rf1] (rf3) {\scriptsize RF};
\node [rectangle,draw,fill=corecolor,minimum width=2.5em,minimum height=1.15em,below=0.1em of rf2] (rf4) {\scriptsize RF};
\node [rectangle,draw,fill=corecolor,minimum width=2.5em,minimum height=1.15em,below=0.1em of rf3] (rf5) {\scriptsize RF};
\node [rectangle,draw,fill=corecolor,minimum width=2.5em,xshift=1.3em,minimum height=1.15em,below=0.1em of rf4] (rf5) {\scriptsize RF};

\node [above=0em of pr3.south] {Raw filters};

\node [rectangle,draw,fill=staticcolor,minimum width=3.5em,minimum height=\blockHeight,right=\blockdist of pr3] (dma) {DMA};



\coordinate (mid) at ($(net.west)!0.5!(dma.east)$);
\coordinate (mid_n) at (mid|-pr3.north);
\node[above=\blockdist of pr3,anchor=mid,inner sep=0] (pl_l) {Programmable Logic};
\coordinate (r_ne) at (dma.east |- pl_l.north);
\coordinate (r_nw) at (net.west |- pl_l.north);
\coordinate (r_sw) at (net.west |- net.south);
\coordinate (r_se) at (dma.east |- net.south);


\begin{pgfonlayer}{background}
    \path (r_ne)+(0.5em,0.5em) node (box_ne) {};
    \path (r_sw)+(-0.5em,-0.5em) node (box_sw) {};
    \path[rectangle,draw,fill=acccolor]
        (box_ne) rectangle (box_sw) ;
\end{pgfonlayer}


\coordinate (r_ne) at (dma.east |- dma.north);
\path (r_se)+(1.5em,0em) node (con0) {};
\node [draw,fill=ramcolor,dashed,minimum width=6.5em,minimum height=2em,anchor=south west] at (con0) (pcie) {PCIe};

\node [draw,fill=ramcolor,minimum width=6.5em,minimum height=2em,above=0.5em of pcie] (ram) {RAM};

\path (box_ne)+(1em,0.5em) node (cpu_pos) {};
\node [rectangle,draw,fill=cpucolor,minimum width=6.5em,minimum height=3.0em, anchor= south west,above=1.0em of ram](ps) {};
\node [rectangle,draw,fill=corecolor,minimum width=2.5em,minimum height=1.5em,above left=0.1em of ps.south] (core0) {\scriptsize Core};
\node [rectangle,draw,fill=corecolor,minimum width=2.5em,minimum height=1.5em,above right=0.1em of ps.south] (core1) {\scriptsize Core};
\node [below=0em of ps.north] {CPU System};



\draw[big_arrow] (net) to (pr3);
\draw[big_arrow_d] (pr3) to (dma);
\draw[mid_arrow] (ps) -| (dma);
\draw[big_arrow_d] (dma.east|-ram) to (ram);
\draw[big_arrow] (dma.east|-pcie) to (pcie);
\draw[big_arrow_d] (ram) to (ram|-ps.south);


\end{tikzpicture}
    \caption{Overview of the proposed architecture for filtering raw byte streams.}
    \label{fig:rpu_overview}
    \vspace{-0.3cm}
\end{figure}
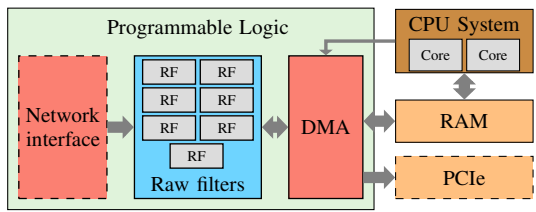


The \glspl{rf} have been evaluated on a Xilinx ZC706 Zynq-7000 SoC. 
\Cref{fig:rpu_overview} depicts the architecture of our system, based on \cite{mci/Becher2019}, consisting of a tightly coupled \gls{ps} and \gls{pl}.
We built 7 parallel pipelined \glspl{rf}, each processing one byte per cycle, leading to a theoretical bandwidth of 1.4 GB/s at a clock rate of 200 MHz. 

In our experiment, 44~MB of inflated JSON data from the RiotBench was preloaded into the RAM and transferred to the RFs using DMA. 
The results, containing only the match signals, were again written back to the RAM using DMA.
In the experiment a data rate of 1.33~GB/s could be achieved, which is sufficient to process incoming data from a 10 GBit/s network interface at line rate.
Accordingly, our system would also be suitable for passing the ingress data from a network interface directly to the RFs.
The filtered data can then be transferred with the DMA into the RAM, in order to process the data directly on the on-chip ARM CPU.
Such a setup could be used, for example, as an IoT gateway that directly performs pre-processing of the received data.
Alternatively, a SmartNIC can be implemented by forwarding the filtered data to a host CPU via PCIe.
Thus, the \glspl{rf} can significantly increase the achieved data rate or relieve the host CPU without a risk of performance degradation.
Since the presented \glspl{rf} require only a small amount of resources, even more \glspl{rf} can be used to process multiple data streams in parallel. 
Furthermore, the programmable logic can be reconfigured, allowing the \glspl{rf} to be replaced when a new query is to be executed.

\section{Conclusion and future work}
\label{sec:Conclusion and future work}
Raw filters have the potential to relieve the CPU workload and create a potential speedup for CPU-bound stream processing applications. 
Even for I/O-bound applications, it may be possible to free up CPU cycles. 
Unlike CPU-based solutions, our versatile primitives allow us to filter a variety of different data sets.
Moreover, the \glspl{rf} can be configured in such a way that the best possible \gls{fpr} is achieved for a given resource requirement.

Currently, the \glspl{rf} are created manually by brute force searching for Pareto points. 
Since this is too time-consuming for an automatic generation of \glspl{rf}, meta heuristics such as evolutionary algorithms can be used in the future. 
Instead of evaluating each design point for the complete dataset, we want to explore sampling methods that can potentially speed up the process without a large increase in the \gls{fpr}.
Furthermore, there are options to further optimize the presented filter primitives which need to be investigated. 
This can be done, for example, by omitting substrings in the string search, or by adjusting the bounds of value range filters, potentially allowing further resource savings without a large increase in \textit{false-positives}.




\footnotesize{
    \bibliography{bibliography}
}

\end{document}